\documentstyle[11pt]{article}
\newcommand{\beq}{\begin{equation}}
\newcommand{\eeq}{\end{equation}}
\newcommand{\beqa}{\begin{eqnarray}}
\newcommand{\eeqa}{\end{eqnarray}}

\title{Postmodern Technicolor}
\author{Thomas Appelquist\\
Department of Physics, Yale University, New Haven, CT 06511
\\ \\
John Terning\\
Department of Physics, University of California, Berkeley, CA  94720
\\  \\
L.C.R. Wijewardhana \\
Department of Physics, University of Cincinnati, Cincinnati,
OH 45221}
\date{June 16, 1997}

\begin{document}
\setlength{\baselineskip}{24pt}
\maketitle
\begin{picture}(0,0)(0,0)
\put(280,335){YCTP-P7-97}
\put(280,320){UCB-PTH-97/12}
\put(280,305){UCTP-9-97}
\end{picture}
\vspace{-36pt}

\begin{abstract}
Using new insights into strongly coupled gauge theories arising from analytic
calculations and lattice simulations, we explore a framework for technicolor
model building that relies on a non-trivial infrared fixed
point, and an essential role for QCD.
Interestingly, the models lead to a simple relation between the electroweak
scale and the QCD confinement scale, and to the possible existence of exotic
leptoquarks with masses of several hundred GeV.
\end{abstract}

Electroweak symmetry breaking and fermion mass generation are still open
problems in particle physics.  The minimal standard model can account for all
current experiments, but, with its light Higgs boson, is technically unnatural.  
An early proposal for avoiding this problem was given in the form of a new 
scaled-up, QCD-like interaction: technicolor (TC).  Problems with accounting
for the charm
and strange quark masses without  flavor changing neutral currents ruled out
such a simple possibility and led to the development of modern TC
theories \cite{walking}, known as walking TC:  theories with vanishing or small
ultraviolet $\beta$ functions.

Precision electroweak measurements have
shown, however, that even walking  theories may be inadequate since they
seem to predict \cite{S} too large a value for the $S$ parameter. The
measurement of the top quark
mass provides an additional problem, since it seems difficult to produce a
large enough top quark
mass without a very small scale for the additional (e.g. extended
technicolor (ETC)) interactions necessary
to couple the technifermion condensate to the top
quark.  Such interactions violate weak isospin symmetry, since they
must also produce a small bottom quark mass. But without suppression by a
relatively large scale, they lead to problems with the $T$
parameter (a.k.a. $\Delta \rho_* = \alpha T$).

In this letter, we make use of  some recent observations about $SU(N)$ gauge
theories to explore a technicolor framework that provides a potential
solution to these problems. The most important of the observations is that
$SU(N)$ theories can exhibit an infrared (IR) fixed point \cite{FP} for a certain range
of flavors\footnote{Such fixed points have also been found in supersymmetric
\cite{susy} gauge theories.}. This behavior has been examined further through analytic
studies \cite{QCDFP,moreFP} and lattice
simulations \cite{lattice}.   An IR fixed point, which  naturally
incorporates walking, is an essential ingredient in this framework for
postmodern technicolor theories (PTC). The use of an IR fixed point to
implement walking was considered by Lane and Ramana as part of their study of
multiscale technicolor models \cite{laneramana}. An important difference,
however, is that in the framework discussed here, the  technicolor fixed point
coupling is not strong enough by itself to break the electroweak symmetry. The
addition of QCD is necessary, and this has interesting consequences.

Consider an $SU(N)$ gauge theory with $N_f$ flavors. At two loops it has a
non-trivial IR fixed point for a range of $N_f$; the coupling at the fixed
point
is given by the ratio of the first two coefficients in the
$\beta$ function:
\beq
\alpha_* = - \,{{ b}\over {c}}~,
\label{alpha*}
\eeq
where
\beq
b = {{1}\over{6 \pi}} \left( 11 N - 2 N_f\right)
\label{b}
\eeq
\beq
c = {{1}\over {24  \pi^2}}
\left(34 N^2 - 10  N N_f - 3{{N^2 -1}\over{N}} N_f\right)~.
\label{c}
\eeq

The two-loop solution $\alpha(q)$ to the renormalization group equation can be
written in the form
\beq
{{1}\over{\alpha(q)}} = b \log\left({{q}\over{\Lambda}}\right) +
{{1}\over{\alpha_*}}
\log\left({{\alpha(q)}\over{\alpha_*-\alpha(q)}}\right),
\eeq
where $\Lambda$ is an intrinsic scale, and where it has been assumed that
$\alpha(q) \leq \alpha_*$. With this choice of scale, $\alpha(\Lambda) \simeq
0.782 \, \alpha_*$.
Then for $q^2 \gg \Lambda^2$ the running coupling displays the usual
perturbative behavior
$\alpha(q) \approx  1/ [b~log(q/\Lambda)]$,
while for $q^2 < \Lambda^2$ it approaches the fixed point $\alpha_*$.

Analytic studies have indicated that there is a critical value, $\alpha_c$, of
the gauge coupling such that if $\alpha > \alpha_c$ then chiral
symmetry breaking takes place. In the presence of an IR fixed point
$\alpha_*$, the chiral transition takes place when  $\alpha_*$ reaches
$\alpha_c$, which happens when $N_f $ decreases to a certain critical value
$N_f^c$  \cite{QCDFP}.
 Below $N_f^c$ the fermions are massive, the fixed point is only
approximate, and we expect confinement to set in at momentum scales on the
order of the fermion mass.
The two-loop CJT potential
\cite{CJT} relates the critical coupling to the quadratic Casimir of the
fermion representation, and for fundamental representations gives:
\beq
\alpha_c \equiv {{ \pi }\over{3 \, C_2(R)}}= {{2 \pi \,
N}\over{3\left(N^2-1\right)}}~.
\label{alphac}
\eeq
This leads to the estimate \cite{QCDFP}:
\beq
N_f^c = N \left({{100N^2 -66}\over{25 N^2 -15}}\right).
\label{Ncrit}
\eeq

The  order parameter (the dynamical fermion mass) vanishes continuously as $N_f
\rightarrow N_f^c$ from below \cite{QCDFP}. Thus for $N_f $ just below
$N_f^c$,
the dynamical mass of the technifermion will be small compared to the intrinsic
scale $\Lambda$. In such a near-critical theory,  the dynamical mass (as a
function of Euclidean momentum $p$)
falls off approximately like $1/ p$ up to
scales of order $\Lambda$, rather than the perturbative (QCD-like) $1/p^2$.
This is due to the fact that the coupling is near the (approximate) IR fixed
point, so it evolves slowly (walks) in this regime \cite{QCDFP}.

Even though no obvious small parameter is involved, an estimate of the next
order term in the loop expansion describing the chiral symmetry
breaking indicates that the
correction is relatively small (less than $20\%$) \cite{QCDFP}.  Similarly, the
next order term (in the $\overline{\rm MS}$ scheme) in the $\beta$ function is
also
about $20\%$ of the first or second order terms when $\alpha_*
\simeq \alpha_c$. We will assume here that the estimates described
above provide an approximate description of the IR fixed point and the chiral
phase transition.

To implement these ideas we take the PTC
group to be $SU(2)_{PTC}$, and assume
that there are four electroweak doublets  of
technifermions ($N_f = 8$).  The motivation for four doublets is
that this corresponds to one complete family of technifermions, i.e.
techniquarks
$Q = (U, D)$ and technileptons $L = (N, E)$.  We assume two technicolors in
order to keep $S$ as small as possible.
Given this theory, Eq.~(\ref{Ncrit}) predicts $N_f^c \simeq 7.9$,
so we might expect  a
conformal IR fixed point, leaving the electroweak symmetry intact.

Note however that the difference between $\alpha_*$ and $\alpha_c$ in this
theory is of order 0.1, which is also the strength of the QCD coupling at the
weak scale. Thus it is quite possible that the inclusion of QCD is
sufficient to produce chiral symmetry breaking in the color singlet channel for
the techniquarks. Suppose, for
example, that there is some  unification scale above $\Lambda$ for all gauge
couplings.  As we evolve the couplings to the IR, the QCD and PTC
interactions grow.  Below $\Lambda$ the PTC coupling  approaches its
fixed point value $\alpha_*$, which by itself is slightly sub-critical for
chiral symmetry breaking. Eventually
the QCD coupling grows to be of order 0.1.
The scale at which the combined interactions reach criticality determines the
techniquark dynamical mass.

At momentum scales below the techniquark mass, there
is no longer an approximate IR fixed point, and the PTC coupling
grows.  We therefore expect that at a somewhat smaller scale, chiral
symmetry breaking will also occur for the technileptons. The magnitude of the
splitting is difficult to estimate reliably since it involves the running of
the TC coupling in the near-critical regime, and will therefore be very
sensitive to the $20\%$ uncertainties discussed above. In this paper, we will
assume that the splitting is sizeable.
The electroweak scale will then be set dominantly by the techniquark dynamical
mass,
with a smaller contribution from the technileptons.
At  momentum scales below these masses, the QCD coupling increases from its
value of approximately $0.1$, eventually reaching  confinement strength at
$\Lambda_{QCD} $.

To be more explicit, we first note that the inclusion of the QCD coupling
$\alpha_s$
modifies the PTC fixed point behavior. There is an additional term,
$({2}/{\pi^2})  \alpha^2 \alpha_s$, in the
two-loop PTC $\beta$ function. Therefore the IR fixed point is only a
quasi-fixed
point (for small $\alpha_s$ and thus for momentum scales large compared to
$\Lambda_{QCD}$) given
by
\beq
\hat \alpha_* = -{{b}\over{c}} +{{2 \alpha_s}\over{c \, \pi^2}}~.
\label{quasi}
\eeq

 We next note that QCD effects also modify\footnote{This is equivalent to the
big MAC analysis of ref.~\cite{AppelTern}.} the two-loop CJT criterion for
chiral symmetry breaking to be
\beq
{{16}\over{9}} \,\alpha_s(\mu) +\alpha(\mu) = \alpha_c = {{4 \pi}\over{9}},
\label{criticality}
\eeq
where $\mu$ is the techniquark chiral symmetry breaking (electroweak) scale.
If the PTC coupling is near its IR quasi-fixed point, then $\alpha(\mu)
\approx \hat \alpha_* $.

At lower momentum scales, the technifermions decouple and the evolution of
$\alpha_s$ is determined by its one-loop renormalization group equation, with
the ultraviolet boundary condition given by solving Eqs. (\ref{quasi}) and
(\ref{criticality}). Thus for $\Lambda \gg \mu > \Lambda_{QCD}$, the
electroweak scale
$\mu$ is related to $\Lambda_{QCD}$ by
\beq
\mu = \Lambda_{QCD} \, \exp\left( {{44}\over{45 \,b_{QCD} (\alpha_c
-\alpha_*)}} \right) ~,
\label{mu}
\eeq
where (from Eq. (\ref{b})) $b_{QCD} = 21/6 \pi$.
Thus in this theory, the electroweak scale can be
computed in terms of the QCD confinement scale.  Of course this result is
exponentially sensitive  to small errors in the estimate of $\alpha_c$, and a
numerically reliable calculation of $\mu$ may require non-perturbative
methods.  Eq. (\ref{mu}) predicts $\log(\mu/\Lambda_{QCD})$  to within
approximately
20\% of the experimental value.

We now consider the effect of the near critical PTC dynamics on  electroweak
physics, first discussing vacuum alignment and the $S$ parameter. $S$ was
estimated for a
one-family $SU(2)_{TC}$ model in Ref.~\cite{revenge}. However, there it was
assumed that the TC dynamics was essentially QCD-like, and that strong ETC
effects explicitly broke the global flavor symmetry from $SU(16)$ (since there
are 8 flavors, but there is no distinction between ${\bf 2}$'s and
$\overline{{\bf 2}}$'s in $SU(2)$) to $SU(8)_L \times SU(8)_R$, which was then
spontaneously broken to $SU(8)_V$. This assumption was important for producing
the correct vacuum
alignment \cite{vacuum}.

Here we are not relying on assumptions
about ETC dynamics, but on the combined effect of QCD and near-critical PTC.
The effect of  QCD is two-fold. First, the $SU(16)$ symmetry is explicitly
broken
down to $SU(3)_c \times SU(2) \times SU(2) \times SU(4)
\times U(1)_ Q \times U(1)$, where $U(1)_Q$ corresponds to techniquark number.
This symmetry is then
spontaneously broken, first to $SU(3)_c \times SU(2)_V \times SU(4) \times
U(1)_Q$
by the techniquark condensate, and finally to
$SU(3)_c \times SU(2)_V \times Sp(4) \times U(1)_Q$ by the technilepton
condensate.

QCD is essential in ensuring that the chiral symmetry breaking for the
techniquarks produces the correct vacuum alignment \cite{vacuum}.
Assuming that spectral density functions of an $SU(2)$ technicolor
theory are similar to those of QCD, it has been shown
that  the chiral symmetry
breaking of the technileptons ($SU(4) \rightarrow Sp(4)$) will break
electromagnetism rather than $SU(2)_L$ \cite{vacuum}.  Here, however, the
chiral symmetry breaking scale for the technileptons, ${\cal O}(4 \pi f_L)$,
is below that of the techniquarks,  ${\cal O}(4 \pi f_Q)$. If the splitting of
scales
is sizeable enough, then at the scale
where the technileptons condense, the techniquarks will have already
broken $SU(2)_L \times U(1)_Y$ down to $U(1)_{em}$. It can then be argued that
electromagnetism remains
unbroken since the
$SU(2)_L$ gauge boson contribution to the
vacuum energy is cutoff in the IR by the $W$ and $Z$ masses. In general,
the analysis is complicated by the fact that the spectral density functions for
a PTC theory may bear little resemblance to their QCD analogs.
In what follows, we will assume that correct vacuum alignment is achieved.

The techniquark symmetry breaking produces the three Goldstone bosons
required for the $W$ and $Z$ masses and an additional techni-axion common to
one-family models. The technilepton symmetry breaking
$SU(4) \rightarrow Sp(4)$ produces five PNGB's. Three have electroweak quantum
numbers and mix with the techniquark Goldstones. We anticipate that
the combination orthogonal to
the longitudinal gauge bosons, composed primarily of technileptons, will
receive
a large mass from new, high energy
interactions (unspecified here) that explicitly break the $SU(4)$ symmetry. The
other two technilepton PNGB's are $SU(2)_L$ and QCD singlets,
and remain massless without the new $SU(4)$-breaking interactions. The
techni-axion must also rely on new, high energy interactions to provide a mass.
In this
case, the interactions must explicitly break the $U(1)$ symmetry.

The approximate $SU(16)$ symmetry of the model implies that there are
110 additional (colored) scalars, which would be conventional PNGB's if
$SU(2)_{PTC}$ were strong enough by itself to spontaneously break
$SU(16)$. Their masses could then be computed perturbatively in
$\alpha_s$  \cite{vacuum} and would be of  of
${\cal O}(\sqrt{\alpha_s} 4 \pi f_Q)$. Here, with QCD required to
help with the breaking, the computation of their masses is subtle,
since as $\Lambda_{\rm QCD} \rightarrow 0$ there is no chiral symmetry
breaking, and therefore no Nambu-Goldstone bosons. It is plausible, however, to
assume that  if we consider
fluctuations around the broken vacuum, perturbative QCD effects
still lead to masses  of
${\cal O}(\sqrt{\alpha_s} 4 \pi f_Q)$ for these scalars. They could also
receive even larger masses  from the new, high energy interactions discussed
above.

Of the 110 colored scalars, 56 have already been considered in
Ref.~\cite{revenge}, where they were true PNGB's. Their contribution to $S$,
along with that of the colorless PNGB's discussed above was estimated to be no
larger than $0.6$, with the contribution becoming smaller as their
masses approach $4 \pi f_Q$. Of the 54 new colored scalars, the $SU(3)_c$
triplet leptoquarks
and diquarks produce a one-loop \cite{S} contribution to $S$ which we estimate
to be  less than ${\cal O}(0.1)$.
Turning to contributions to $S$ arising at scales $4 \pi f_Q$ and above, it was
noted in Ref.~\cite{revenge}
that if there is a sizeable mass splitting between techniquarks and
technileptons, as could arise in the present theory, and a splitting
between the technielectron
and technineutrino, then the technileptons could also give a negative
contribution to $S$. While the model  described so far has no mechanism
for the splitting of the technielectron and technineutrino, the smaller
technilepton
masses will be more sensitive than the
techniquark masses to the new, high energy interactions necessary to give mass
to the quarks and leptons.

Putting all this together, our crude
estimates, including those of Ref.~\cite{revenge}, suggests that there is a
significant range of parameter space
where the full contribution to $S$ in this model
may lie below the $95 \%$ confidence experimental upper limit of approximately
$0.17$. Of course, a truly reliable estimate of $S$ is not yet available in a
non-QCD like technicolor theory, and this is especially true of this model with
its complex spectrum of techniparticles.

We next consider the isospin breaking effects resulting from the
interactions (unspecified here\footnote{Such interactions could be generated by
a light
composite scalar which may form in the breaking of a chiral gauge theory down
to a theory with an IR fixed point \cite{lightscalar}.}) that produce the
coupling of the top quark to
the techniquark condensate and hence the top quark mass.  To accomplish this
we first need an estimate of the Goldstone boson decay constant  $f_Q$ in the
techniquark sector.  If the techniquarks and technileptons were degenerate,
then using the relation $f_L^2 + 3 f_Q^2 = v^2$ we would have $f_Q = f_L
= v/2 \simeq 123$ GeV.  In the present model the techniquarks are heavier
than the technileptons, and for purposes of numerical estimates we will simply
take
$f_Q =2 f_L$ (i.e. $f_Q \simeq 136$ GeV).

Recent quenched lattice results
\cite{quarkmass} suggest that the average of the up and down quark masses,
$\hat m$, is smaller than folklore assumes: $\hat m \approx 3.6 MeV$.
Calculations with dynamical quarks find even smaller values for $\hat m$
\cite{quarkmass}.  Using the well known relation between the quark condensate
$\langle \bar\psi \psi \rangle$, $\hat m$, and the pion mass, this leads to a
new
estimate of the QCD condensate:
\beq
{{\langle \bar\psi \psi \rangle}\over{f_\pi^3}} = {{ m_\pi^2}\over{2 \hat m
f_\pi}} \approx 27 \approx 8 \pi.
\eeq

Using this relation to estimate the techniquark condensate, we have for the
top quark mass
\beq
m_t \approx {{27 f_Q^3}\over{\Lambda_t^2}} {{\Lambda_t}\over{f_Q}} ~,
\label{mt}
\eeq
where $\Lambda_t$ is the scale\footnote{The scale $\Lambda_t$ can in principle 
be below the scale where PTC is embedded in a larger gauge group.} 
of the physics that induces the top quark
coupling\footnote{Depending on how this coupling is produced
there may be an additional factor of $N_c$ in the condensate factor.}
to the condensate. The factor $\Lambda_t/
f_Q$ accounts for the high-energy enhancement effects of IR fixed point
(walking) dynamics \cite{walking}.   Using $m_t = 175$ GeV, we find $\Lambda_t
\simeq
2.9$ TeV. In  making this estimate, we have assumed that $\Lambda \gg
\Lambda_t$ so that $\alpha$ stays very close to $\alpha_*$, and therefore the
dynamical techniquark mass falls only like $1/p$, up to momenta of order
$\Lambda_t$. If  this is not the case, then a smaller value  of $\Lambda_t$
will be necessary. We have also taken the walking to start at the scale $f_Q$,
which is an additional uncertainty in the calculation.

In generic models, we expect that these isospin violating interactions which
produce $m_t$ will
also induce isospin violating, effective four-techniquark interactions, which
contribute to $\Delta \rho_*$. To lowest order in the interaction this so
called ``direct" contribution can be estimated to be:
 \beq
\Delta \rho_*^{d} \approx {{\left( 3
f_Q^2
\right)^2}\over {v^2 \Lambda_t^2}} ,
\eeq
which less than 0.5 \% as long as
$\Lambda_t  > 3.2$ TeV. In Ref.~\cite{revenge}, it was  pointed out that there
are also
potentially negative contributions to $\Delta \rho_*$ arising from PNGB's.
To higher order in the isospin violating interactions, there will be
additional contributions to $\Delta \rho_*$. These include effects that arise
from the mass difference between the $U$ and $D$ techniquarks. In order that
they also remain beneath the experimental bound, the $U$ and $D$ must be nearly
degenerate \cite{AES}.

To conclude, we have explored a new framework for technicolor model building
which relies
on the near-critical behavior of a theory with an approximate IR fixed
point. Since QCD effects are crucial for rendering the fixed point only
approximate and
producing electroweak symmetry breaking, the weak scale is predicted in terms
of the QCD scale.  The combination of near-critical PTC and QCD can
significantly split
the techniquarks and technileptons, and provide a framework for reducing the
value of
$S$ \cite{revenge}. Whether the electroweak gauge
symmetry is broken in the correct pattern, and whether the predicted weak scale
and the deviations from the minimal standard model agree with experiment will
require further study.

\noindent \medskip\centerline{\bf Acknowledgements}
\vskip 0.15 truein
We  thank  N. Arkani-Hamed, C. Carone, R.S. Chivukula, K. Lane, H. Murayama,
and S. Selipsky for helpful discussions.
This work was supported in part by  the Department of Energy under
contracts \#DE-FG02-92ER40704 and  \#DE-FG-02-84ER40153. J.T. is supported by
the National Science Foundation under grant PHY-95-14797.

\vskip 0.15 truein

%\vfill\eject
%\noindent \medskip\centerline{\bf Figure Captions}
%\vskip 0.15 truein

\end{document}